\documentclass{conm-p-l}

\copyrightinfo{2017}{}

\usepackage{graphicx}
\usepackage{subfigure}

\usepackage{amssymb,amsmath,amsthm,amscd}

\theoremstyle{definition}

\theoremstyle{remark}

\numberwithin{equation}{section}

\newcommand{\tu}{\tilde{u}}

\newcommand{\hk}{\hat{k}}

\newcommand{\e}{\epsilon}

\newcommand{\ga}{\gamma}

\newcommand{\Dl}{\Delta}
\renewcommand{\th}{\theta}
\newcommand{\ra}{\rightarrow}
\newcommand{\al}{\alpha}

\newcommand{\sg}{\sigma}

\newcommand{\pa}{\partial}

\newcommand{\La}{\Lambda}

\newcommand{\om}{\omega}

\newcommand{\na}{\nabla}
\newcommand{\lag}{\langle}
\newcommand{\rag}{\rangle}

\newcommand{\non}{\nonumber}

\newcommand{\hu}{\hat{u}}

\begin{document}

\title[Short term unpredictability]{Short term unpredictability of high Reynolds number turbulence --- rough dependence on initial data}

\author{Z. C. Feng}
\address{Z. C. Feng, Department of Mechanical and Aerospace Engineering, 
University of Missouri, Columbia, MO 65211}
\email{fengf@missouri.edu}

\author{Y. Charles Li}
\address{Y. Charles Li, Department of Mathematics, University of Missouri, 
Columbia, MO 65211, USA}
\email{liyan@missouri.edu}
\urladdr{http://faculty.missouri.edu/~liyan}

\curraddr{}
\thanks{}



\date{}

\dedicatory{}


\begin{abstract}
Short term unpredictability is discovered numerically for high Reynolds number fluid flows under periodic boundary conditions. Furthermore, the abundance of the short term unpredictability is also discovered. These discoveries support our theory that fully developed turbulence is constantly driven by such short term unpredictability.
\end{abstract}

\maketitle

\section{Introduction}

Turbulence as an open problem has two aspects: turbulence engineering and turbulence physics \cite{Li13}. Turbulence engineering deals with how to describe turbulence in engineering. Turbulence physics deals with the physical mechanism of turbulence. In pursuit of understanding of turbulence physics, recently we proposed the theory that fully developed turbulence is caused by short term unpredictability due to rough dependence upon initial data, while (often) transient turbulence at moderate Reynolds number is caused by chaos (long term unpredictability) \cite{Li14}. The main goal of this article is to 
demonstrate the short term unpredictability via numerical simulations. According to our analytical theory \cite{Li14}, perturbations in turbulence can amplify in time according to $e^{\sg \sqrt{Re} \sqrt{t} + \sg_1 t}$ where $\sg_1 = \frac{\sqrt{2e}}{2} \sg$ and $\sg$ depends only on the base solutions on which the perturbations are introduced, and $Re$ is the Reynolds number. When the time is small, the first term in the exponent dominates, and this term can cause the amplification to be superfast when the Reynolds number is large. By the time $t \sim Re$, the two terms in the exponent are about equal. After the time $t \sim Re$, the second term dominates, and this term is the classical Liapunov exponent that causes chaos (long term unpredictability). Thus the time $t \sim Re$ is the temporal separation point between short term unpredictability and long term unpredictability. When the Reynolds number is large, long before the separation point $t \sim Re$, the first term in the exponent already amplifies the perturbation to substantial size so that the nonlinear effect takes over, and the second term does not get a chance to dominate. Thus fully developed turbulence is dominated by such short term unpredictability.  When the Reynolds number is moderate, both terms in the exponent have a chance to dominate, and the corresponding (often) transient turbulence is dominated by chaos in long term. 

Long term unpredictability has been well understood. The main feature of chaos is long 
term unpredictability led by sensitive dependence on initial data. On the other hand, short term unpredictability 
is led by rough dependence on initial data. If the solutions are non-differentiable in their initial data, as in 
the case of Euler equations of fluids \cite{Inc12}, then any small initial perturbation will be amplified to 
order $O(1)$ instantly. Such a short term unpredictability is very different from the long term unpredictability of 
chaos. Such a short term unpredictability is closer to total randomness than the  long term unpredictability of 
chaos.  Nevertheless, such a short term unpredictability is still not total randomness, for instance the solutions
of Euler equations of fluids are still continuous in their initial data, and the conserved quantities do not vary too much under perturbations. Such a short term unpredictability leads to a 
peculiar process that is very close to a random process but still constrained. When the Reynolds number is 
moderate, dynamics of Navier-Stokes equations is quite far away from that of Euler equations. Turbulence 
at such a stage is often transient, and bears clear resemblance to finite dimensional chaos \cite{VK11} \cite{KE12} \cite{GS16} \cite{LK15} \cite{DSH10} \cite{Vis07} \cite{KUV12} \cite{KK01}. One can name 
such turbulence as chaos in Navier-stokes equations. Such turbulent solutions are differentiable in 
their initial data (at least during the known time interval of existence), and the derivatives of the solutions in initial data have moderate norms. 
When the Reynolds number is very high, dynamics of Navier-Stokes equations is getting closer to that of Euler 
equations. High Reynolds number turbulence is fully developed, and has no resemblance to finite dimensional chaos. Such turbulent solutions are still differentiable in 
their initial data (at least during the known time interval of existence), but the derivatives of the solutions in initial data have huge norms in the order of $e^{\sg \sqrt{Re} \sqrt{t} + \sg_1 t}$ mentioned above which represent the growth rate of the perturbations. Thus initial perturbations are amplified super fast even in short time. We believe that this causes the abrupt nature in the development of high Reynolds number turbulence. Since perturbations constantly exist, there are constantly such super fast amplifications of perturbations which lead to the persistence nature of high Reynolds number turbulence (so-called fully developed turbulence) in contrast to the transient nature of moderate 
Reynolds number turbulence.

In terms of phase space dynamics of dynamical systems, when the Reynolds number is very high, fully developed
turbulence is not the result of a strange attractor, rather a result of super fast amplifications of ever present perturbations. Strange 
attractor is a long time object, while the development of such violent turbulence is of short time. Such fully developed
turbulence is maintained by constantly super fast perturbation amplifications. When the Reynolds number is set to 
infinity, the perturbation amplification rate is infinity. So the dynamics of Euler equations is very close to a random process.
In contrast, chaos in finite dimensional conservative systems often manifests itself as the so-called stochastic 
layers. Dynamics inside the stochastic layers has the long term sensitive dependence on initial data.
When the Reynolds number is moderate, viscous diffusive term in Navier-Stokes equations is stronger, perturbation
amplification rate is moderate. At this stage, turbulence is basically chaos in Navier-Stokes equations \cite{VK11} \cite{KE12}  \cite{GS16} \cite{LK15} \cite{DSH10} \cite{Vis07}
\cite{KUV12} \cite{KK01}. In some cases, strange attractor can be observed \cite{VK11}. 

The article is organized as follows: In section 2, we briefly review Liapunov exponent and chaos. In section 3, we briefly review analytical results on rough dependence. In section 4, we are going to shed 
new light on the classical hydrodynamic instability theory from a new perspective. In section 5, 
2D numerical demonstration on rough dependence is presented. In section 6, 
3D numerical demonstration on rough dependence is presented. Section 7 is the conclusion.

\section{Chaos --- sensitive dependence on initial data} 

There are many ways to characterize chaos, and one necessary ingredient of every 
characterization is ``sensitive dependence on initial data''. For solutions that exhibit sensitive dependence on initial data, their initial small perturbations are usually amplified exponentially (with an exponent named Liapunov exponent), and it takes time for the perturbations 
to amplify to substantial size (say order $O(1)$ relative to the small initial 
perturbations). If $\e$ is the initial small perturbation size, and $\sg$ is the Liapunov exponent, 
then the time for the perturbation to reach order $O(1)$ is about 
\[
\frac{1}{\sg} \ln \frac{1}{\e}.
\]
The Liapunov exponent $\sg$ is a long term object defined by
\[
\sg = \lim_{t \ra +\infty} \lim_{d u_0 \ra 0} \frac{1}{t} \ln \frac{ \| d u(t) \|}{\| d u_0 \|},
\]
where $d u_0$ is the initial perturbation, and $\| \ \|$ is certain norm. 
Positive Liapunov exponent usually is a good indicator of chaos (even though the matter can be 
tricky sometimes \cite{LK07}). In the phase space of the dynamics, when the Liapunov exponent 
is positive, initially nearby orbits diverge exponentially with the exponential rate being the Liapunov 
exponent. If these orbits are bounded in the phase space, then it is intuitively natural to expect the 
dynamics being chaotic. 

There are of course other ways for solutions of deterministic systems to be ``irregular'' than that of  
chaotic solutions. Next we will describe another way: rough dependence on initial data.

\section{High Reynolds number turbulence --- rough dependence on initial data} 

Turbulent motion of fluids is modeled by the so-called Navier-Stokes equations.
The phase space of the dynamics of Navier-Stokes equations is infinite dimensional. 
The well known such a phase space is the Sobolev space of divergence free fields, 
$H^n(\mathbb{R}^d)$ ($d=2,3$) which 
contains functions that are square-integrable and so are their derivatives up to $n$-th order.
When $n > \frac{d}{2} +1$ ($d=2,3$), for any initial condition in such a phase space, 
it is known  \cite{Kat72} \cite{Kat75}  that there is a (short) time $T>0$ depending on the norm of 
the initial condition, such that the corresponding solution (orbit) of 
Navier-Stokes equations (and Euler equations) exists on [$0,T$].  Such an orbit is continuous in time 
$t$ and its initial condition. As the Reynolds number $Re \ra \infty$, the solution of 
Navier-Stokes equations converges to that of the Euler equation. In two dimensions ($d=2$), the 
existence time $T$ is infinite, while in three dimensions ($d=3$), global existence is still an open problem.
The above claims apply also to spatially periodic domain $\mathbb{T}^d$ in stead of $\mathbb{R}^d$.

One can define a solution map in the phase space by mapping  the initial condition to 
the solution's value at time $t$. The solution map for Euler equations ($d=2,3$) is continuous,
but nowhere uniformly continuous, and more importantly
nowhere differentiable \cite{Inc12}. Then it is natural to expect that the norm of the derivative of the solution 
map for  Navier-Stokes equations approaches infinity as the Reynolds number approaches infinity.
Under Euler dynamics, any small perturbation of the initial condition can potentially reach substantial amount 
instantly. It is natural to expect that under high Reynolds number Navier-Stokes dynamics, 
small perturbation of the initial condition can potentially reach substantial amount in a very short time
(the larger Reynolds number, the shorter). We call this phenomenon ``rough dependence on initial 
data''. Such rough dependence on initial data naturally leads to the violent fully developed 
turbulence as observed in experiments. One can try to estimate the size of the derivative of the solution 
map for  Navier-Stokes equations. The  Navier-Stokes equations are given by
\begin{eqnarray}
& & u_t - \frac{1}{Re} \Dl u  = - \na p - u\cdot \na u , \label{NS} \\
& & \na \cdot u = 0 , 
\end{eqnarray}
where $u$ is the $d$-dimensional fluid velocity ($d=2,3$), $p$ is the fluid pressure, and 
$Re$ is the Reynolds number. Setting the Reynolds number to infinity $Re = \infty$, the 
Navier-Stokes equations (\ref{NS}) reduces to the Euler equations
\begin{eqnarray}
& & u_t  = - \na p - u\cdot \na u , \label{E} \\
& & \na \cdot u = 0 .
\end{eqnarray}
For any $u \in H^n(\mathbb{R}^d)$, there 
is a neighborhood $B$ and a short time $T>0$, such that for any $v \in B$ there exists a unique 
solution to the Navier-Stokes equations (\ref{NS}) in $C^0([0,T]; H^n(\mathbb{R}^d))$. As $Re 
\ra \infty$, this solution converges to that of the Euler equations (\ref{E}) in the same space. For any 
$t \in [0, T]$, let $S^t$ be the solution map:
\begin{equation}
S^t \  :  \   B  \mapsto H^n(\mathbb{R}^d), \  S^t (u(0)) = u(t), \label{SM} 
\end{equation}
i.e. the solution map maps the initial condition to the solution's value at time $t$. The solution map 
is continuous for both Navier-Stokes equations (\ref{NS}) and Euler equations (\ref{E}) \cite{Kat72} 
\cite{Kat75}. A recent result of Inci \cite{Inc12} shows that for Euler equations (\ref{E}) the solution map is 
nowhere differentiable. Even though the derivative of the solution map for Navier-Stokes equations (\ref{NS})
exists, it is natural to conjecture that the norm of the derivative of the solution 
map approaches infinity as the Reynolds number approaches infinity.  The following upper bound was 
obtained in \cite{Li14}.
\begin{equation}
\| DS^t(u(0)) \| = \sup_{d u(0)} \frac{\| d u(t) \|}{\| d u(0) \|} \leq e^{\sg \sqrt{Re} \sqrt{t} \ + \ \sg_1 t}, \label{UB} 
\end{equation}
where $d u(0)$ is any initial perturbation of $u(0)$, and 
\[
\sg = \frac{8}{\sqrt{2e}} \max_{\tau \in [0,T]} \| u(\tau )\|_n, \ \ \sg_1 = \frac{\sqrt{2e}}{2} \sg .
\]
The above bound also 
applies to spatially periodic domain $\mathbb{T}^d$ in stead of $\mathbb{R}^d$.
The main aim of this article is to numerically demonstrate that in fully developed turbulence, perturbations 
amplify according to the growth rate given by the right hand side of (\ref{UB}).

\section{Classical hydrodynamic instability --- directional derivative}

Classical hydrodynamic instability theory mainly focuses on the so-called linear instability of steady fluid 
flows.  We can think that the linear instability theory is based on Taylor expansion of the solution map for 
Navier-Stokes equations (\ref{NS}). Let $u^*$ be the steady flow (a fixed point in the phase space), 
$v_0$ be its initial perturbation, and $u^* + v(t)$ be the solution to the Navier-Stokes equations (\ref{NS}) 
with the initial condition  $u^* + v_0$.  According to Taylor expansion,
\[
v(t) = dv(t) + d^2v(t) + \cdots ,
\]
where $dv(t)$ is the first differential in $u^*+v_0$ of the solution map at the steady flow $u^*$, similarly for $d^2v(t)$ etc..
Under the Euler dynamics, this expansion fails since the first differential does not exist \cite{Inc12}. Under the Navier-Stokes dynamics, this expansion is valid, and the 
first differential satisfies the differential form
\begin{eqnarray}
& & dv_t - \frac{1}{Re} \Dl dv  = - \na dp - dv\cdot \na u^* - u^*\cdot \na dv , \label{dNS} \\
& & \na \cdot dv = 0 ,
\end{eqnarray}
where $dp$ is the pressure differential. The linear instability refers to the instability of the differential form 
(\ref{dNS}). In most cases studied, the steady flow $u^*$ depends on only one spatial variable $y$ (the so-called 
channel flow). This permits the following type solutions to the differential form,
\begin{equation}
dv(t) = \exp \{ i (\sg t + k_1 x + k_3 z) \} V(y) , \label{FM}
\end{equation}
where ($x,y,z$) are the spatial coordinates, $\sg$ is a complex parameter, and ($k_1,k_3$) are real parameters.
One can view (\ref{FM}) as a single Fourier mode out of the Fourier transform of $dv(t)$. In the phase space of the 
dynamics, (\ref{FM}) is a directional differential with the specific direction specified by the ($k_1,k_3$) Fourier mode.
$V(y)$ satisfies the well-known Orr-Sommerfeld equation (Rayleigh equation in the inviscid case $Re = \infty$).
Even though the first differential $dv(t)$ does not exist in the inviscid case ((\ref{dNS}) with $Re = \infty$), the 
directional differential (\ref{FM}) can exist with $V(y)$ solving the Rayleigh equation. Thus, the linear stability/instability predicted by the Rayleigh equation only represents a directional linear 
stability/instability of the Euler dynamics while the full first differential of the Euler dynamics 
does not exist. The classical 
hydrodynamic instability theory heavily focuses on the studies of the Rayleigh equation. The directional linear instability derived from Rayleigh equation often imply linear instability in 
Orr-Sommerfeld equation \cite{LL11}. Nevertheless, linear instability due to unstable eigenvalues cannot capture the dominant  linear instability of super fast growth. For a detailed evaluation on the rigorous mathematical foundation of linear hydrodynamic stability theory, see \cite{Li18}.

\section{2D numerical simulations on rough dependence on initial data} 

In this section, we will demonstrate numerically the super fast amplification of 
perturbations to the solutions of 2D Navier-Stokes equations. In particular, we shall demonstrate that such super fast amplification of perturbations is ubiquitous. Microscopically, Navier-Stokes equations model fluid flows well. Thus, the super fast amplification phenomenon in Navier-Stokes equations also reflects the same phenomenon in physical fluid flows.         

\subsection{A fundamental problem in the numerical simulations} 

\begin{figure}[ht]
\centering
\includegraphics[width=4.5in,height=4.5in]{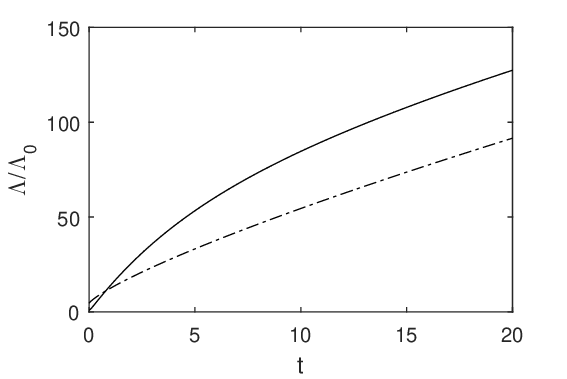}
\caption{This figure clearly illustrates the fundamental obstacle in numerical simulation on the norm of the solution 
map's derivative as $t \ra 0^+$. The dash curve represents the analytically obtained lower bound (\ref{ar}) on the norm of the
directional derivative of a family of explicit solutions, where $\ga = 0.6$ and $\sg =27.5$ are chosen. The solid curve represents the numerical simulation on the norm of the
directional derivative of the same family of explicit solutions. One can see clearly that near $t=0$, the rigorous lower bound 
is violated. In particular, the dash curve has infinite derivative at $t=0$, while the solid curve has finite derivative.}
\label{LBN}
\end{figure}

First we numerically simulate an explicit example \cite{Li15} to test the numerical performance. Consider the 2D 
Navier-Stokes equations
\begin{equation}
\pa_t u + u \cdot \na u = - \na p + \frac{1}{Re} \Dl u, \  \na \cdot u = 0 , \label{2DNS}
\end{equation}
under periodic boundary condition with period domain [$0, 2\pi$] $\times$ [$0, 2\pi$], where $u=(u_1,u_2)$ is the 
velocity, $p$ is pressure, and the spatial coordinate is denoted by $x=(x_1,x_2)$. A simple solution to the 2D 
Navier-Stokes equations (\ref{2DNS}) is 
\begin{equation}
u_1 = \sum_{n=1}^\infty \frac{1}{n^{3+\ga }} e^{-\frac{n^2}{Re}t}\sin [n (x_2-\sg t)] , \  u_2 = \sg , \label{NSsim}
\end{equation}
where $\frac{1}{2} < \ga \leq 1$, and $\sg$ is a real parameter. By varying $\sg$, we get a variation direction of 
the initial condition,
\[
du_1(0) = 0 , \quad du_2(0) = d \sg , 
\]
which leads to the variation of the solution ($du_1(t),du_2(t)$). Let
\[
\La = \| (du_1(t),du_2(t)) \|_{H^3}, \ \La_0 = \| (du_1(0),du_2(0)) \|_{H^3}, 
\]
then one has the analytical result \cite{Li15}:
\begin{equation}
\frac{\La}{\La_0} \geq \left ( 1 +\left [ \frac{1}{\sqrt{2e}} t^{\ga} \left (\frac{\sqrt{t} \sqrt{Re}}{2\sqrt{2}}\right )^{1-\ga } \right ]^2 \right )^{1/2}. \label{ar}
\end{equation}
This lower bound is obtained by keeping only the fastest growing mode given by 
\begin{equation}
n = \bigg [ \sqrt{\frac{Re}{2t} }\bigg ] , \  \  \bigg (\text{the integer part of } \sqrt{\frac{Re}{2t}} \bigg ) .
\label{mgm}
\end{equation}
Notice that as $t \ra 0^+$, the time derivative of the lower bound approaches positive infinity due to the fractional power of $t$. 
That is, the lower bound curve is tangent to the vertical axis at $t=0$. 
As $t \ra 0^+$, the fastest growing mode (\ref{mgm}) $n \ra +\infty$. Thus a numerical simulation will never capture the fastest growing mode as $t \ra 0^+$ no matter how many Fourier modes are kept in the numerical simulation. This demonstrates a fundamental problem in numerical simulations. When we numerically simulate the quantity $\frac{\La}{\La_0}$ (\ref{ar}), we obtained the solid curve in Figure \ref{LBN}. Notice that as $t \ra 0^+$, the numerical solid curve gets below the dash lower bound curve (violating the lower 
bound nature). The numerical solid curve has a finite time derivation at $t=0$, and does not capture the infinite derivative nature at $t=0$.

\subsection{Fixed base solution and different perturbations} 

\begin{figure}[ht] 
\centering
\subfigure[$Re=1000$]{\includegraphics[width=2.3in,height=2.3in]{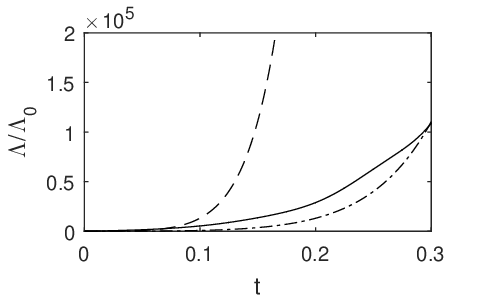}}
\subfigure[$Re=1000$]{\includegraphics[width=2.3in,height=2.3in]{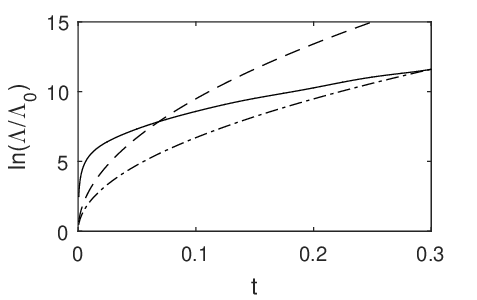}}
\subfigure[$Re=100000$]{\includegraphics[width=2.3in,height=2.3in]{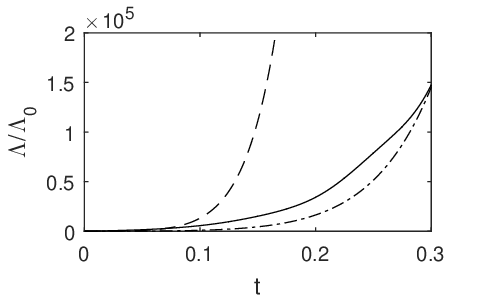}}
\subfigure[$Re=100000$]{\includegraphics[width=2.3in,height=2.3in]{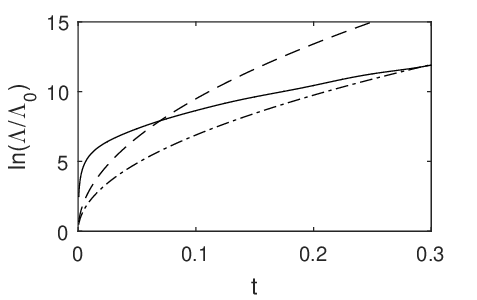}}
\caption{The solid curve is the numerical result of the super fast growth of perturbations with initial 
condition (\ref{PSI})-(\ref{IS1}) where $\La (t) = \| du(t) \|_{H^3}$. The lower fitting dashed curve is $e^{21.2 \sqrt{t}}$ when 
$Re =1000$ and $e^{21.7 \sqrt{t}}$ when $Re =100000$. The closest fitting dashed curve is $e^{30 \sqrt{t}}$ when 
$Re =1000$ and $Re =100000$.}
\label{CM}
\end{figure}

\begin{figure}[ht] 
\centering
\subfigure[$Re=1000, k_1=k_2=1$]{\includegraphics[width=2.3in,height=2.3in]{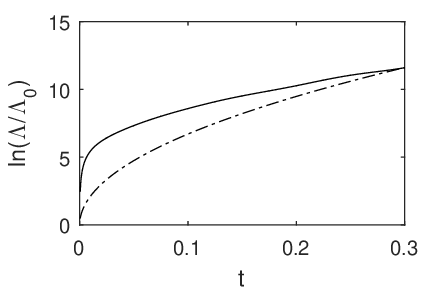}}
\subfigure[$Re=1000, k_1=k_2=2$]{\includegraphics[width=2.3in,height=2.3in]{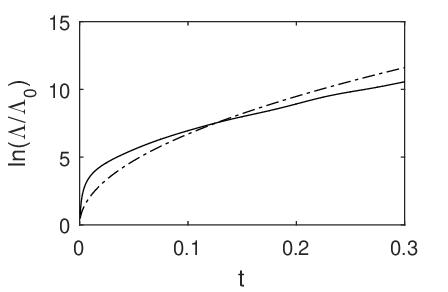}}
\subfigure[$Re=1000, k_1=k_2=3$]{\includegraphics[width=2.3in,height=2.3in]{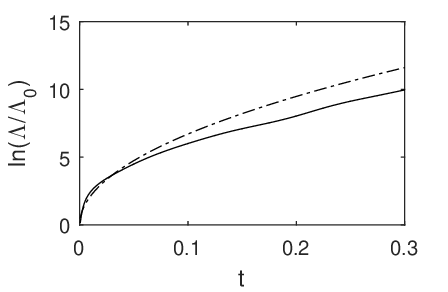}}
\subfigure[$Re=1000, k_1=k_2=4$]{\includegraphics[width=2.3in,height=2.3in]{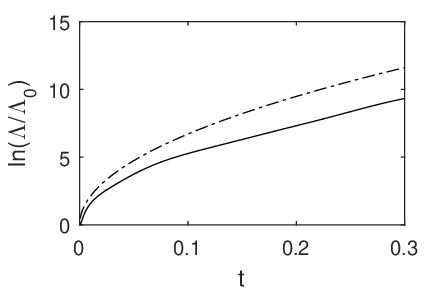}}
\caption{The solid curve is the numerical result of the super fast growth of perturbations with initial 
condition (\ref{PSI}) with different ($k_1,k_2$) where $\La (t) = \| du(t) \|_{H^3}$. The fitting dashed curve is $e^{21.2 \sqrt{t}}$.}
\label{GR1}
\end{figure}

\begin{figure}[ht] 
\centering
\subfigure[$Re=1000, k_1=k_2=5$]{\includegraphics[width=2.3in,height=2.3in]{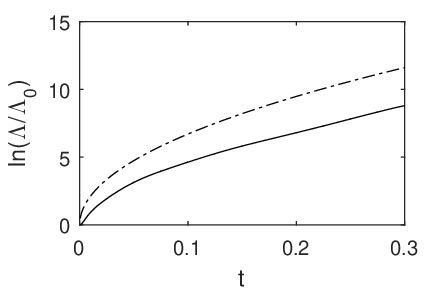}}
\subfigure[$Re=1000, k_1=k_2=6$]{\includegraphics[width=2.3in,height=2.3in]{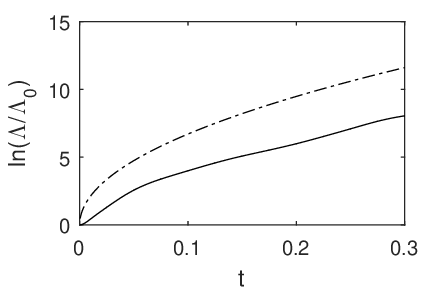}}
\subfigure[$Re=1000, k_1=k_2=7$]{\includegraphics[width=2.3in,height=2.3in]{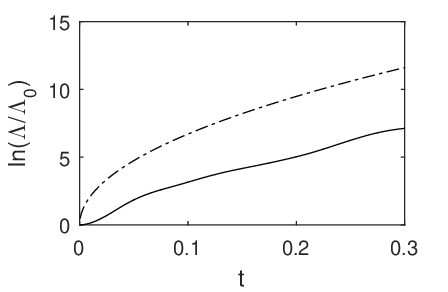}}
\subfigure[$Re=1000, k_1=k_2=8$]{\includegraphics[width=2.3in,height=2.3in]{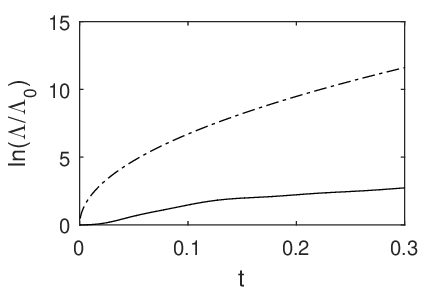}}
\caption{The solid curve is the numerical result of the super fast growth of perturbations with initial 
condition (\ref{PSI}) with different ($k_1,k_2$) where $\La (t) = \| du(t) \|_{H^3}$. The fitting dashed curve is $e^{21.2 \sqrt{t}}$.}
\label{GR2}
\end{figure}

We will numerically simulate the 2D Navier-Stokes equations under periodic boundary condition (\ref{2DNS}). We have two goals here: First, we want to realize the super fast 
amplification of perturbations. Second, we want to show that such super fast 
amplification of perturbations is abundant among perturbations. For the two goals, we 
shall choose the initial conditions of the base solution and the perturbations, to be of the form of single Fourier modes. Since the perturbation equations are linear, perturbation solutions generated from such single 
Fourier modes form a base of superposition. For the 
base solution, we choose the initial condition
\begin{equation}
u_1(0) = -8\sin (9x_1) \sin (8x_2), u_2(0) = -9\cos (9x_1) \cos (8x_2),
\label{BSI}
\end{equation}
Starting from this initial condition, we solve (\ref{2DNS}) numerically to generate the base solution. The perturbation $du$ 
based upon a base solution $u$ solves the linearized 2D Navier-Stokes equations,
\begin{equation}
\pa_t du + u \cdot \na du + du \cdot \na u= - \na dp + \frac{1}{Re} \Dl du, \  \na \cdot du = 0 , \label{2DLNS}
\end{equation}
under the same periodic boundary condition as in (\ref{2DNS}). 
Since (\ref{2DLNS}) is linear, we can choose single Fourier modes as the initial conditions of the perturbations, 
\begin{equation}
du_1(0) = -0.1k_2\sin (k_1x_1) \sin (k_2x_2), du_2(0) = -0.1k_1\cos (k_1x_1) \cos (k_2x_2).
\label{PSI}
\end{equation}
Figure \ref{CM} shows the super fast growth of the perturbation when
\begin{equation}
k_1 =1,k_2=1, Re = 1000 \text{ and } Re =100000, \label{IS1}
\end{equation} 
where the time step for the numerical simulation is $\Dl t = 0.0005$. 
We use the notation  
\begin{equation}
\La (t) = \| du(t) \|_{H^3} .  \label{land}
\end{equation}
Then the norm of the derivative of the solution map at the base solution $u(t)$ is given by
\begin{equation}
\| DS^t(u(0)) \| = \sup_{du(0)} \frac{\La (t)}{\La(0)}. 
\label{NR}
\end{equation}
Notice that for any fixed $t$, the supremum is taken with respect to all initial perturbation $du(0)$. If one initial 
perturbation leads to a perturbation that is near the supremum for some $t$, it may not be near the supremum for other $t$.
The norm of the derivative $\| DS^t(u(0)) \|$ has an upper bound given by (\ref{UB}). We 
anticipate that the nature of the square root of time in the exponent of the upper 
bound (\ref{UB}) can be realized by an individual perturbation, while the nature of 
the square root of the Reynolds number can only be realized by the supremum over a lot of perturbations. For any particular perturbation,
the viscous effect is negligible when the Reynolds number is relatively large. On the other
hand, a generic perturbation in physics contains all``basic perturbation directions", and 
the fastest growing direction will quickly dominate the amplification. In fact, the fastest growing direction may change in time. Due to numerical obstacles such as that demonstrated 
in Figure \ref{LBN}, numerical simulations as $t \ra 0^+$ are not quite reliable. That is 
the reason that our $\frac{1}{\ln t} \ln \ln \frac{\La}{\La_0}$ numerical simulations do not converge to a constant. Nevertheless, in appropriate time interval, we are confident that our numerical simulations clearly show super fast growth (faster than exponential growth) as demonstrated in Figure \ref{CM}. Increasing 
the wave number ($k_1,k_2$), the perturbation's growth rate decreases as shown in Figures \ref{GR1} - \ref{GR2}. When the wave number of the initial perturbation is larger, the viscous effect is more significant. Our conclusion is that the super fast growth (rough dependence) is abundant among perturbations in the sense that generic perturbations contain all Fourier modes, and low Fourier 
modes display the super fast growth. Next we shall study the abundance of the super fast growth 
among base solutions, that is, whether or not there are abundant base solutions of which the perturbations have super fast growth.


\subsection{Fixed perturbation and different base solutions} 

Our goal here is to show that the super fast amplification of perturbations is abundant among base solutions. For this goal, we will again choose the initial conditions of the base solutions and the perturbation, to be of the form of single Fourier modes. 
First we fix the initial condition of the perturbation to be the case of $k_1 =1$ and $k_2=1$ in (\ref{PSI}), and the Reynolds number 
$Re=1000$. Then we simulate different base solutions with initial conditions of the form,
\begin{equation}
u_1(0) = -k_2\sin (k_1x_1) \sin (k_2x_2), \  u_2(0) = -k_1\cos (k_1x_1) \cos (k_2x_2).
\label{BS2}
\end{equation}

\begin{figure}[ht] 
\centering
\subfigure[$Re=1000, k_1=9,k_2=8$]{\includegraphics[width=2.3in,height=2.3in]{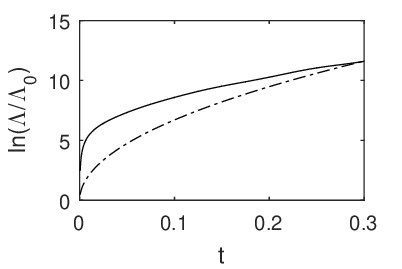}}
\subfigure[$Re=1000, k_1=8,k_2=7$]{\includegraphics[width=2.3in,height=2.3in]{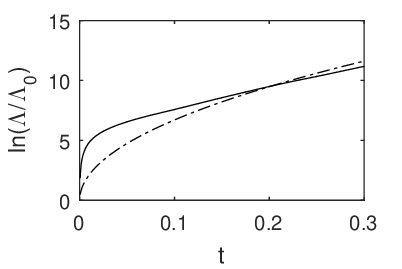}}
\subfigure[$Re=1000, k_1=7,k_2=6$]{\includegraphics[width=2.3in,height=2.3in]{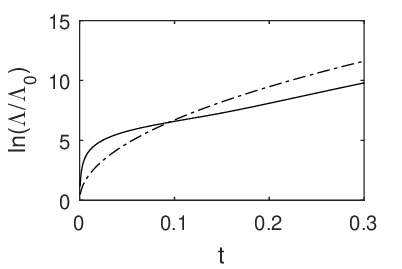}}
\subfigure[$Re=1000, k_1=6,k_2=5$]{\includegraphics[width=2.3in,height=2.3in]{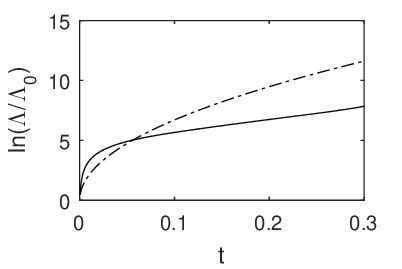}}
\caption{The solid curve is the numerical result of the super fast growth of perturbations with initial 
condition (\ref{PSI}) with $k_1=1$ and $k_2=1$ under different base solutions with initial conditions given by (\ref{BS2}),  where $\La (t) = \| du(t) \|_{H^3}$. The fitting dashed curve is $e^{21.2 \sqrt{t}}$.}
\label{BGR1}
\end{figure}
\begin{figure}[ht] 
\centering
\subfigure[$Re=1000, k_1=5,k_2=4$]{\includegraphics[width=2.3in,height=2.3in]{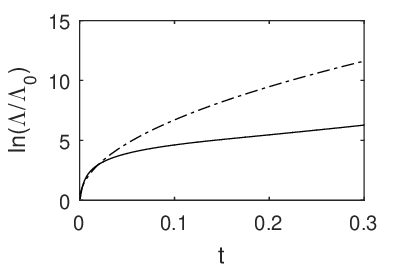}}
\subfigure[$Re=1000, k_1=4,k_2=3$]{\includegraphics[width=2.3in,height=2.3in]{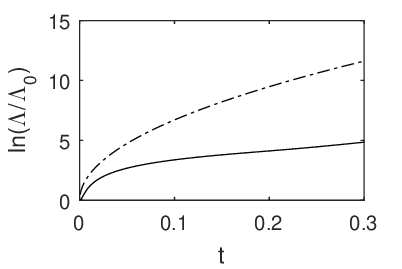}}
\subfigure[$Re=1000, k_1=3,k_2=2$]{\includegraphics[width=2.3in,height=2.3in]{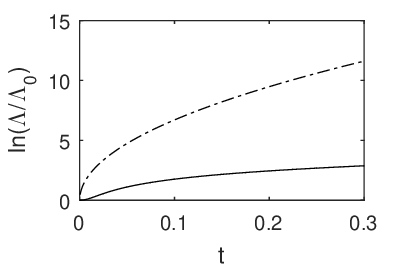}}
\subfigure[$Re=1000, k_1=2,k_2=1$]{\includegraphics[width=2.3in,height=2.3in]{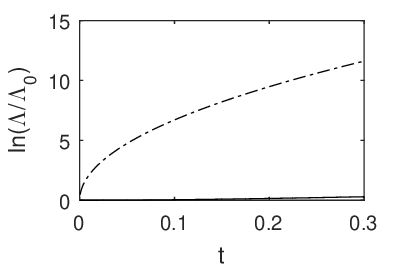}}
\caption{The solid curve is the numerical result of the super fast growth of perturbations with initial 
condition (\ref{PSI}) with $k_1=1$ and $k_2=1$ under different base solutions with initial conditions given by (\ref{BS2}),  where $\La (t) = \| du(t) \|_{H^3}$. The fitting dashed curve is $e^{21.2 \sqrt{t}}$.}
\label{BGR2}
\end{figure}
For different choices of ($k_1,k_2$), the super fast growths are shown in Figures \ref{BGR1}-\ref{BGR2}.
As the wave numbers ($k_1,k_2$) of the base solutions decrease, the super fast growth rates of the perturbation decrease. Together with the result of last subsection, we conclude that higher wave number 
base solutions and lower wave number perturbations correspond to faster super fast growth of the perturbations. Numerical simulations on other base solutions also show super fast growth of the perturbations. Thus super fast growth of the perturbations (rough dependence) is also 
abundant among base solutions. One can then envision that when the Reynolds number is large, 
the super fast growth of the ever present perturbations will cause the abrupt development of turbulence, and is the mechanism that maintains the persistence of turbulence (the so called fully developed turbulence). 

\subsection{Turbulence Regime}

\begin{figure}[ht] 
\centering
\subfigure[Start at $t=0$]{\includegraphics[width=2.3in,height=2.3in]{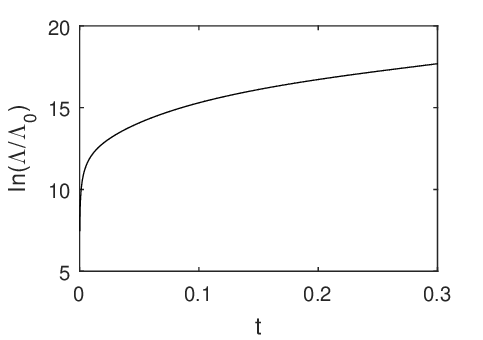}}
\subfigure[Restart at $t=0.03$]{\includegraphics[width=2.3in,height=2.3in]{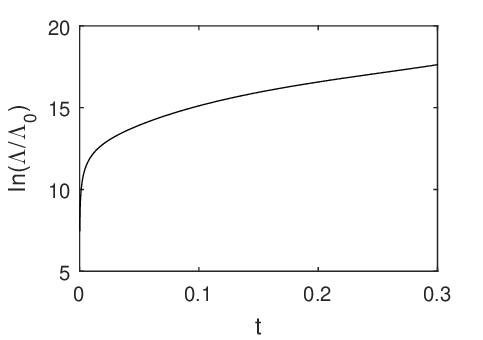}}
\caption{Super fast growths of the perturbations with the same initial 
condition (\ref{pit1}) which is introduced at (a). $t=0$, and (b). $t=0.03$ to the base solution 
with the initial condition (\ref{tbi1})-(\ref{tbi2}),  where $\La (t) = \| du(t) \|_{H^3}$.}
\label{Tur1}
\end{figure}

In this subsection, we shall simulate more realistic situations of base solutions in the turbulence regime. Now start with base solution's initial condition in the form
\begin{eqnarray}
& & u_1(0) = \sum_{0\leq m,n \leq 16} a_{mn} n \sin ( m x_1 + \th_1) \sin ( n x_2 + \th_2) , \label{tbi1} \\
& & u_2(0) = \sum_{0\leq m,n \leq 16} a_{mn} m \cos ( m x_1 + \th_1) \cos ( n x_2 + \th_2) , \label{tbi2}
\end{eqnarray}
where $a_{mn} = 0.01 a$ and $a$ is a random variable with standard Gaussian distribution, and $\th_1$ and 
$\th_2$ are random variables with uniform distribution on [$0,1$]. This type of initial conditions put the flow 
into the turbulence regime. For the perturbation initial condition, we choose
\begin{equation}
du_1(0) = a_1 \sin (x_1) \sin (x_2), du_2(0) = a_1 \cos (x_1) \cos (x_2),
\label{pit1}
\end{equation}
where $a_1$ is a random variable with standard Gaussian distribution (in this single mode case, it does not matter whether or not $a_1$ is random since the perturbation satisfies a linear equation). We choose the Reynolds number $Re = 1000$. First we run the simulation 
for a time period $0 \leq t \leq 0.03$ with time step $0.0005$, the super fast growth of the perturbation is
shown in Fig. \ref{Tur1}(a). Then we restart the base solution from $t=0.03$, i.e. we take the $t=0.03$ flash of 
the base solution as the new initial condition, introduce the same initial perturbation (\ref{pit1}), and run the
simulation for the same time period $0 \leq t \leq 0.03$ with the same time step $0.0005$. The super fast growth of the perturbation is shown in Fig. \ref{Tur1}(b). We conclude that at any moment, an initial perturbation is introduced into turbulence, it immediately goes through a $e^{c \sqrt{t}}$ super fast amplification. We can visualize turbulence as a constant super fast amplification of ever appearing perturbations.

\begin{figure}[ht] 
\centering
\subfigure[$N=16$]{\includegraphics[width=2.3in,height=2.3in]{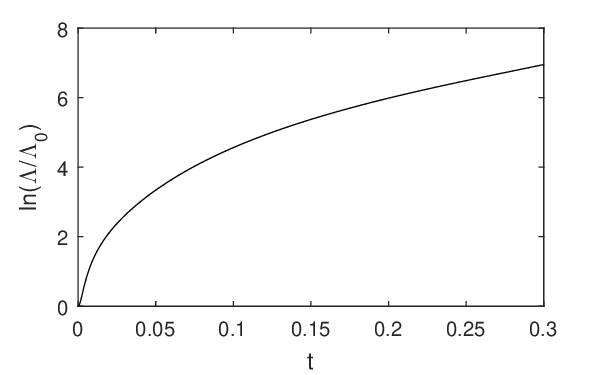}}
\subfigure[$N=8$]{\includegraphics[width=2.3in,height=2.3in]{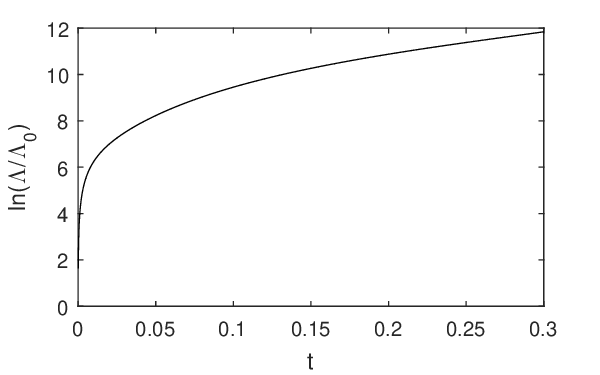}}
\subfigure[$N=4$]{\includegraphics[width=2.3in,height=2.3in]{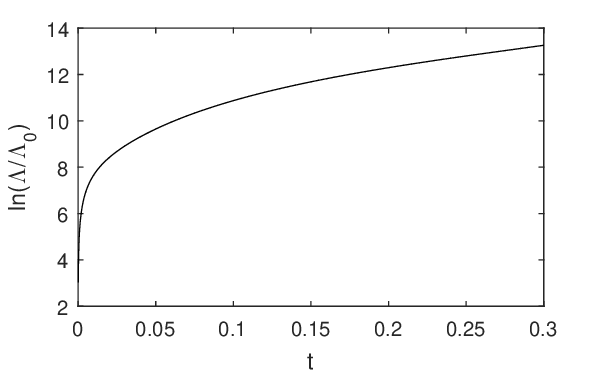}}
\subfigure[$N=2$]{\includegraphics[width=2.3in,height=2.3in]{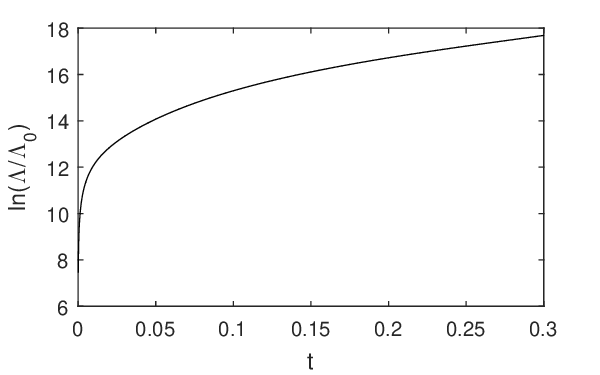}}
\caption{Super fast growths of the perturbations with the initial 
condition (\ref{ptbi1})-(\ref{ptbi2}) when $N=16, 8, 4, 2$, to the base solution 
with the initial condition (\ref{tbi1})-(\ref{tbi2}),  where $\La (t) = \| du(t) \|_{H^3}$.}
\label{Tur2}
\end{figure}

Now we choose more general initial perturbations to the base solution initial condition (\ref{tbi1})-(\ref{tbi2}) as follows
\begin{eqnarray}
& & du_1(0) = \sum_{0\leq m,n \leq N} a_{mn} n \sin ( m x_1 + \th_1) \sin ( n x_2 + \th_2) , \label{ptbi1} \\
& & du_2(0) = \sum_{0\leq m,n \leq N} a_{mn} m \cos ( m x_1 + \th_1) \cos ( n x_2 + \th_2) , \label{ptbi2}
\end{eqnarray}
where the parameters are defined in (\ref{tbi1})-(\ref{tbi2}). When $N=16, 8, 4, 2$, we have the same superfast growth (Fig. \ref{Tur2}), and clearly lower perturbation mode grows faster. Again, since the perturbation equations are linear, each initial individual Fourier mode amplifies independently. 

\subsection{Norm independence of the short term unpredictability}

\begin{figure}[ht] 
\centering
\subfigure[$H^3 \ norm, Re=100$]{\includegraphics[width=2.3in,height=2.3in]{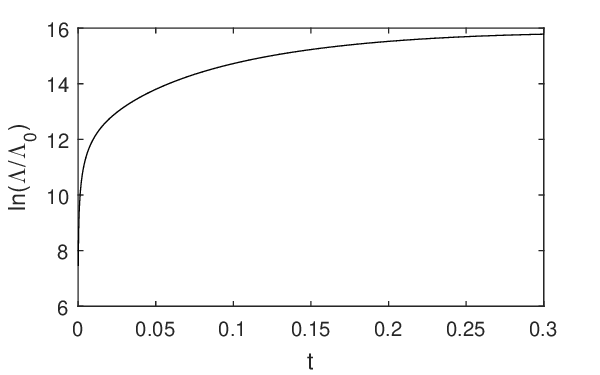}}
\subfigure[$H^3 \ norm, Re=5$]{\includegraphics[width=2.3in,height=2.3in]{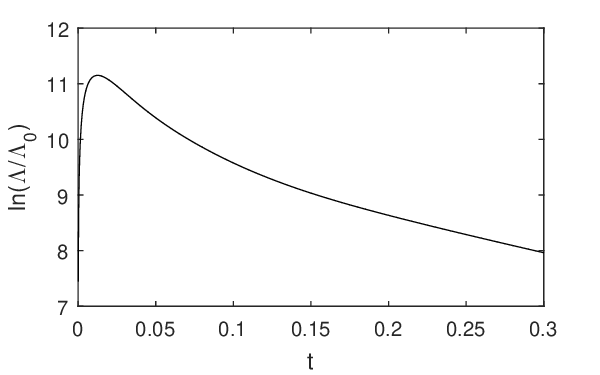}}
\subfigure[$H^0 \ norm, Re=100$]{\includegraphics[width=2.3in,height=2.3in]{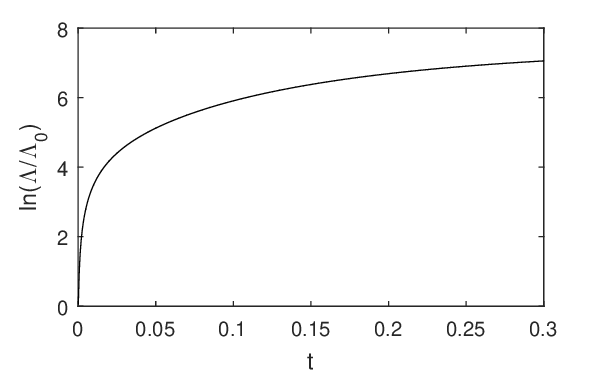}}
\subfigure[$H^0 \ norm, Re=5$]{\includegraphics[width=2.3in,height=2.3in]{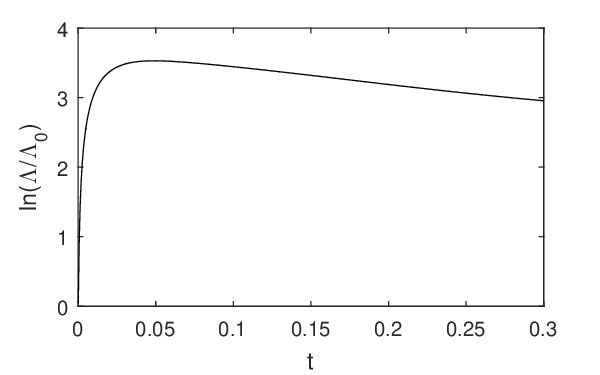}}
\caption{Super fast growths of the perturbations measured in different norms, with the initial 
condition (\ref{ptbi1})-(\ref{ptbi2}) to the base solution with the initial 
condition (\ref{tbi1})-(\ref{tbi2}) ($N=2$),  where $\La (t) = \| du(t) \|_{H^n}$ ($n=0,3$).}
\label{Fnorm}
\end{figure}

One natural question is whether or not the super fast amplification of perturbations is due to the special norm $H^3$. We tested different norms. For all the cases we tested, we observed that the super fast growth (short term unpredictability) nature is independent of the norm used to measure the perturbation. 
Figure \ref{Fnorm} shows the representative cases of $H^0$ and $H^3$ norms. 

In general, norm dependence is a delicate matter \cite{Yud00}. We do not rule out special examples such that the super fast growth depends on norms.

\subsection{An intuition on the superfast amplification of perturbation} 

Take 2D for example, following is an intuition on the superfast amplification of perturbation: In terms of the vorticity variable $\om$, 
a mode $k^b = (k^b_1, k^b_2)$ with amplitude $\om_{k^b}$  of the base solution and a mode $k^p= (k^p_1, k^p_2)$ with 
amplitude $\om_{k^p}$  of the perturbation make a contribution \cite{Li00} 
\[
\frac{1}{2} \left [ |k^b|^{-2} - |k^p|^{-2} \right ] \left | \begin{array}{lr} k^p_1 & k^b_1 \cr 
k^p_2 & k^b_2 \end{array} \right |  \om_{k^p}\om_{k^b}
\]
to the time derivative of the amplitude $\om_{k^p +k^b}$ of the perturbation mode $k^p +k^b$. When $|k^b|$ is much larger 
than $k^p$, the above contribution is usually of the order 
\[
|k^b| | \om_{k^p}|.
\]
This can lead to a super-fast amplification of the perturbation in $H^1$ norm (i.e. vorticity's $L^2$ norm). The amplification of the $H^3$ norm
of the perturbation is even faster. 

This simple intuition only hints a possible amplification. The specific temporary amplification as revealed via numerical 
simulation is of the form $e^{c\sqrt{t}}$.

\section{Reynolds-number dependence of the superfast amplification of perturbations}

\begin{figure}[ht] 
\centering
\subfigure[$\lag d\hu (t), d\tu (t) \rag_{H^3} > 0$]{\includegraphics[width=2.3in,height=2.3in]{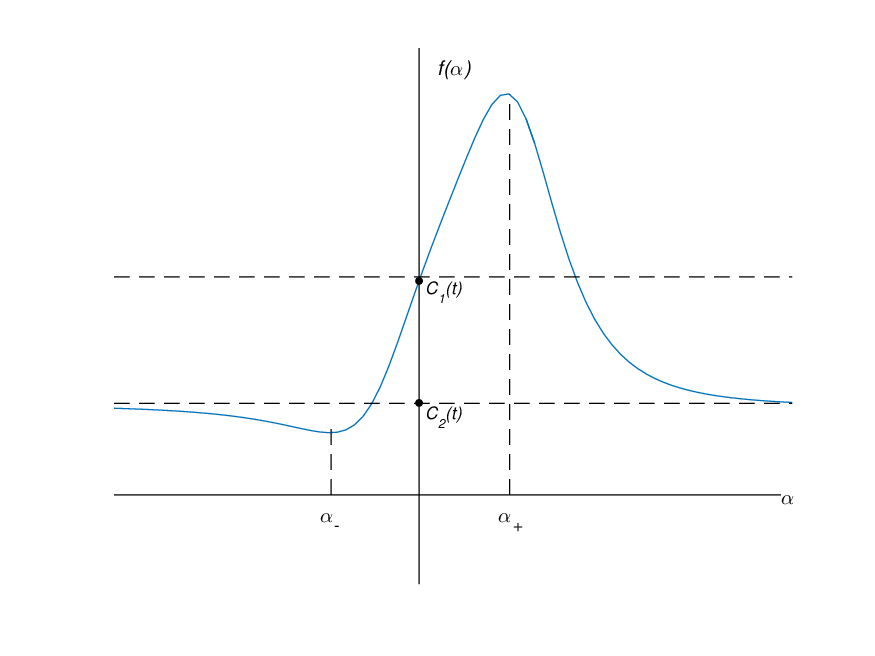}}
\subfigure[$\lag d\hu (t), d\tu (t) \rag_{H^3} < 0$]{\includegraphics[width=2.3in,height=2.3in]{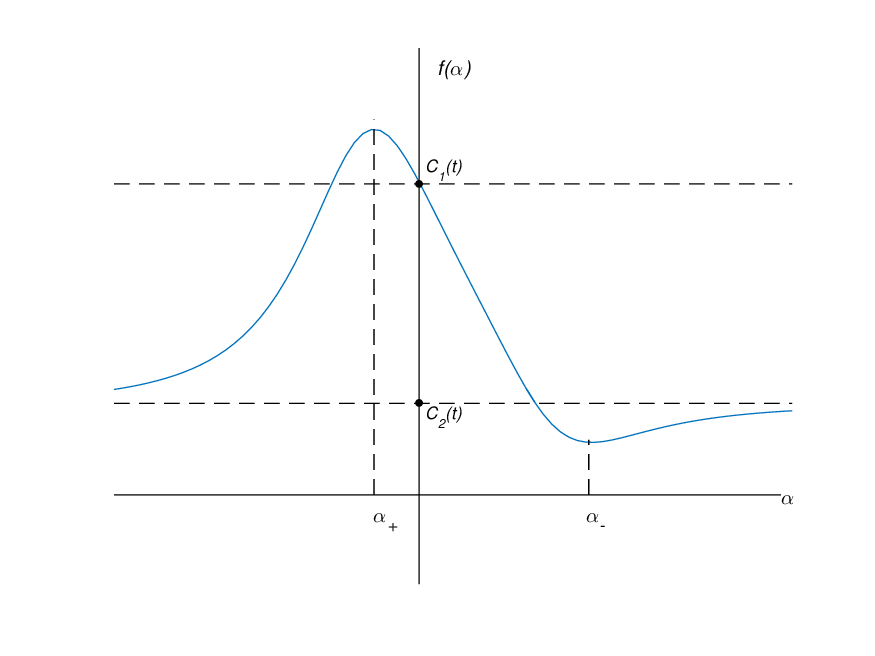}}
\caption{Qualitative feature of $f(\al )$, (a). the case $\lag d\hu (t), d\tu (t) \rag_{H^3} > 0$, and (b). the case $\lag d\hu (t), d\tu (t) \rag_{H^3} < 0$.}
\label{QFf}
\end{figure}

When the initial perturbations are of single modes, lower mode amplifies faster. This does not 
mean that the lowest mode initial perturbation is the maximizer in the definition (\ref{NR}) of the 
norm of the derivative of the solution map, as shown below. Let $d\hu (0)$ and $d\tu (0)$ be two 
single-mode initial perturbations:
\[
\| d\hu (t) \|_{H^3} = \sqrt{C_1(t)} \| d\hu (0) \|_{H^3}, \  \| d\tu (t) \|_{H^3} = \sqrt{C_2 (t)} \| d\tu (0) \|_{H^3}.
\]
For any fixed $t>0$, without loss of generality, assume that $C_1 (t) \geq C_2 (t)$. Consider the 
initial perturbations $d\hu (0) + \al d\tu (0)$, where $\al$ is a real parameter,
\[
\| d\hu (t) + \al d\tu (t) \|_{H^3}^2 =  \| d\hu (t) \|_{H^3}^2 + \al^2 \| d\tu (t) \|_{H^3}^2
+ 2\al \lag d\hu (t), d\tu (t) \rag_{H^3} ,
\]
where 
\begin{eqnarray*}
&& \lag d\hu (t), d\tu (t) \rag_{H^3} = \\
&& \int \sum_{0\leq m+n \leq 3, \ell = 1,2} \left [ \left (
\frac{\pa}{\pa x_1}\right )^m \left (\frac{\pa}{\pa x_2}\right )^n  d\hu_\ell \right ] \left [ \left (
\frac{\pa}{\pa x_1}\right )^m \left (\frac{\pa}{\pa x_2}\right )^n  d\tu_\ell \right ] dx_1 dx_2 .
\end{eqnarray*}
Then
\[
\| d\hu (t) + \al d\tu (t) \|_{H^3}^2 =  C_1(t)\| d\hu (0) \|_{H^3}^2 + \al^2 C_2(t) \| d\tu (0) \|_{H^3}^2
+ 2\al \lag d\hu (t), d\tu (t) \rag_{H^3} .
\]
Since $d\hu (0)$ and $d\tu (0)$ are single modes, $\lag d\hu (0), d\tu (0) \rag_{H^3} = 0$,
\[
\| d\hu (0) + \al d\tu (0) \|_{H^3}^2 =  \| d\hu (0) \|_{H^3}^2 + \al^2 \| d\tu (0) \|_{H^3}^2 .
\]
Next we will show that for some range of the parameter $\al$,
\begin{equation}
\| d\hu (t) + \al d\tu (t) \|_{H^3}^2 >  C_1(t) \| d\hu (0) + \al d\tu (0) \|_{H^3}^2 , 
\label{IEQ}
\end{equation}
that is, $d\hu (0) + \al d\tu (0)$ can amplify faster than both $d\hu (0)$ and $d\tu (0)$. The
inequality (\ref{IEQ}) is equivalent to 
\[
\frac{2}{\al} \lag d\hu (t), d\tu (t) \rag_{H^3} > (C_1(t) - C_2(t)) \| d\tu (0) \|_{H^3}^2 .
\]
As long as $\lag d\hu (t), d\tu (t) \rag_{H^3} \neq 0$, the inequality is satisfied in certain 
range of $\al$ with small enough $|\al |$. Similarly, in another range of $\al$ with large 
enough $|\al |$, such that 
\[
-2\al \lag d\hu (t), d\tu (t) \rag_{H^3} > (C_1(t) - C_2(t)) \| d\hu (0) \|_{H^3}^2 ,
\]
we have
\[
\| d\hu (t) + \al d\tu (t) \|_{H^3}^2 <  C_2(t) \| d\hu (0) + \al d\tu (0) \|_{H^3}^2 ,
\]
that is, $d\hu (0) + \al d\tu (0)$ can amplify slower than both $d\hu (0)$ and $d\tu (0)$.
Let
\[
f(\al ) = \frac{\| d\hu (t) + \al d\tu (t) \|_{H^3}^2}{\| d\hu (0) + \al d\tu (0) \|_{H^3}^2} ,
\]
the qualitative feature of $f(\al )$ is shown in Figure \ref{QFf}, where Figure \ref{QFf}(a) 
corresponds to the case $\lag d\hu (t), d\tu (t) \rag_{H^3} > 0$ and Figure \ref{QFf}(b) 
corresponds to the case $\lag d\hu (t), d\tu (t) \rag_{H^3} < 0$, since 
\[
f'(0) = \frac{2 \lag d\hu (t), d\tu (t) \rag_{H^3}}{\| d\hu (0) \|_{H^3}^2} .
\]

Denote by 
\[
A = \| d\hu (0) \|_{H^3}^2 , \  B = \| d\tu (0) \|_{H^3}^2 , \  P(t) = \lag d\hu (t), d\tu (t) \rag_{H^3} ,
\]
the maximum and minimum points of $f(\al )$ are given by 
\[
\al_{\pm} = \frac{-AB (C_1(t) - C_2(t)) \pm \sqrt{A^2B^2 (C_1(t) - C_2(t))^2 
+4ABP(t)^2}}{2BP(t)} .
\]
In the case $C_1 (t) > C_2 (t)$, when $|\al_+|$ is small, 
\begin{equation}
\al_+ \approx \frac{P(t)}{(C_1(t) - C_2(t))B}, \label{appm}
\end{equation}
when $|\al_-|$ is large, 
\[
\al_- \approx  -\frac{(C_1(t) - C_2(t))A}{P(t)} .
\]
The key point is that if 
\[
\lag d\hu (0), d\tu (0) \rag_{H^3} = 0, \text{ and } \lag d\hu (t), d\tu (t) \rag_{H^3} \neq 0 ,
\]
then there is a range of the parameter $\al$ such that $d\hu (0) + \al d\tu (0)$ can amplify faster than both $d\hu (0)$ and $d\tu (0)$. By iterating the above argument for all the single Fourier 
modes, adding higher and higher Fourier modes leads to faster and faster amplifications. Such 
combinations of more and more Fourier modes amplify closer and closer to the supremum (\ref{NR}). In 
general, the supremum cannot be reached by any specific initial perturbation. The supremum strongly depends on the Reynolds number $Re$. For a specific initial perturbation, the Reynolds-number effect is negligible when the Reynolds number is large enough. In fact, we believe that the square-root 
nature of the Reynolds number in (\ref{UB}) can only be realized by the supremum, and cannot be 
realized by any specific initial perturbation. 

Our argument here shows that in turbulence, neither lower modes nor higher modes rather certain combinations of them grow faster! Such a mechanism manifests in the appearance of turbulence.

\section{3D numerical simulations on rough dependence on initial data} 

In this section, we will demonstrate numerically the super fast amplification of perturbations to the solutions of the 3D Navier-Stokes equations. We will also show that such super fast amplification of perturbations is ubiquitous. 
We numerically simulate the 3D Navier-Stokes equations (\ref{NS}) for the base solutions, and the corresponding 
perturbation equations (the same form with (\ref{2DLNS})) under the periodic boundary condition with period domain 
[$0, 2\pi$] $\times$ [$0, 2\pi$]$\times$ [$0, 2\pi$]. In parallel with the 2D simulations, we have two goals: First we want to realized the super fast amplification of perturbations, and then we want to show that such super fast amplification of perturbations is abundant 
among perturbations and base solutions. Again, for such two goals, we are going to choose the initial conditions of the base solutions and the perturbations, to be of the form of 
single Fourier modes. Since the perturbation equations are linear, such perturbation solutions generated from single Fourier modes form a base of superposition. 

We will start with the initial condition for the base 
solution in the single-mode form
\begin{eqnarray}
u_1(0) &=& \frac{A}{k_1} \cos (k_1 x_1) \sin (k_2 x_2) \sin (k_3 x_3) , \non \\
u_2(0) &=& \frac{A}{k_2} \sin (k_1 x_1) \cos (k_2 x_2) \sin (k_3 x_3) , \label{3dbi} \\
u_3(0) &=& -2\frac{A}{k_3} \sin (k_1 x_1) \sin (k_2 x_2) \cos (k_3 x_3) , \non
\end{eqnarray}
and the initial condition for the perturbation in the single-mode form
\begin{eqnarray}
du_1(0) &=& \frac{a}{\hk_1} \cos (\hk_1 x_1) \sin (\hk_2 x_2) \sin (\hk_3 x_3) , \non \\
du_2(0) &=& \frac{a}{\hk_2} \sin (\hk_1 x_1) \cos (\hk_2 x_2) \sin (\hk_3 x_3) , \label{3dpi} \\
du_3(0) &=& -2\frac{a}{\hk_3} \sin (\hk_1 x_1) \sin (\hk_2 x_2) \cos (\hk_3 x_3). \non
\end{eqnarray}
After we simulate these single-mode cases, we will simulate more realistic situations in the turbulence regime. 

\subsection{Fixed base solution and different perturbations}

To show the abundance of the super fast amplification among perturbations, we choose 
in (\ref{3dbi})-(\ref{3dpi}) that
\begin{equation}
Re = 1000, \ A = 20, \ k_1 =6, \ k_2=5, \ k_3 = 1, \ a=0.1, \label{3dfbe}
\end{equation}
and time step $\Dl t = 0.0001$.    
\begin{figure}[ht] 
\centering
\subfigure[$\hk_1 = 1, \hk_2 = 1, \hk_3 = 1$]{\includegraphics[width=2.3in,height=2.3in]{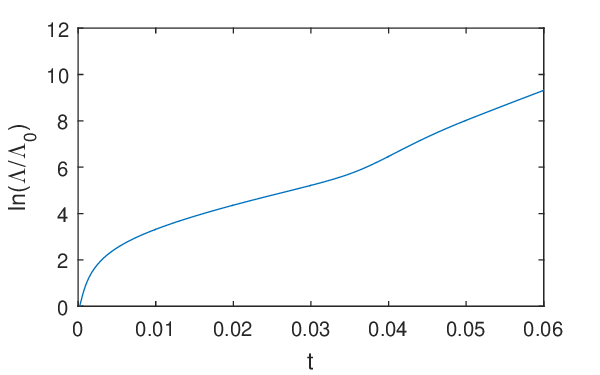}}
\subfigure[$\hk_1 = 2, \hk_2 = 2, \hk_3 = 2$]{\includegraphics[width=2.3in,height=2.3in]{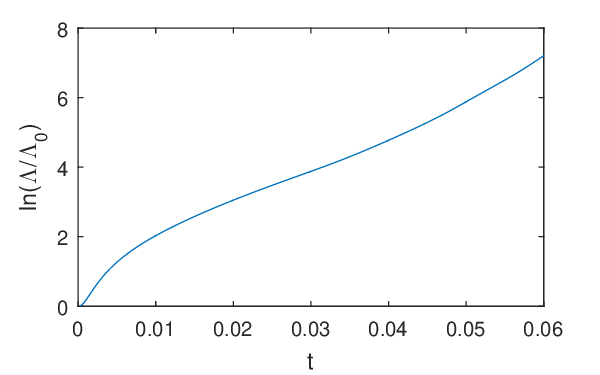}}
\subfigure[$\hk_1 = 3, \hk_2 = 3, \hk_3 = 3$]{\includegraphics[width=2.3in,height=2.3in]{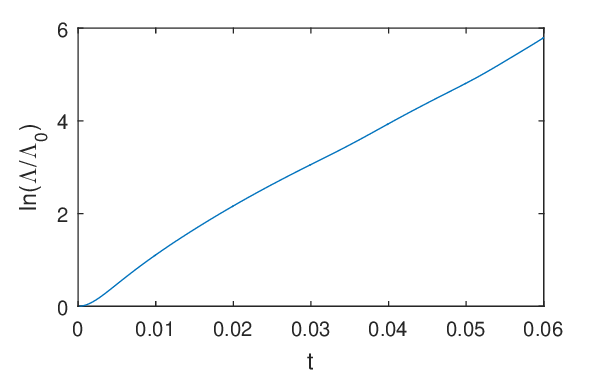}}
\caption{Super fast growth of the perturbations of different modes in 3D with parameters (\ref{3dfbe}).}
\label{3dfbf}
\end{figure}
Figure \ref{3dfbf} shows the super fast growth of the perturbations
of different modes where $\La (t)$ is defined in (\ref{land}).  
We arrive at the same conclusion as in 2D, that is, lower wave number perturbations have faster 
super fast growth, and such super fast growth is abundant among perturbations.

\subsection{Fixed perturbation and different base solutions} 

To show the abundance of the super fast amplification among base solutions, 
we choose in (\ref{3dbi})-(\ref{3dpi}) that
\begin{equation}
Re = 1000, \ A = 20, \ \hk_1 = 1, \ \hk_2 = 1, \ \hk_3 = 1, \ a=0.1, \label{3dfpe}
\end{equation}
and time step $\Dl t = 0.0001$.  
\begin{figure}[ht] 
\centering
\subfigure[$k_1 =6, k_2=5, k_3 = 1$]{\includegraphics[width=2.3in,height=2.3in]{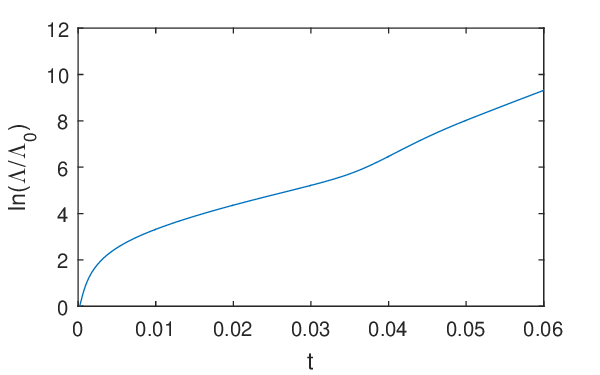}}
\subfigure[$k_1 = 7, k_2 = 6, k_3 = 2$]{\includegraphics[width=2.3in,height=2.3in]{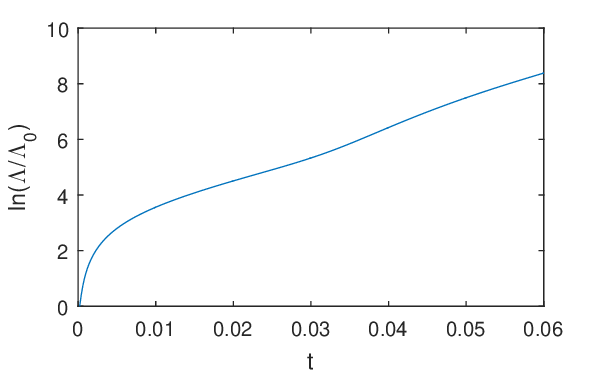}}
\subfigure[$k_1 = 8, k_2 = 7, k_3 = 3$]{\includegraphics[width=2.3in,height=2.3in]{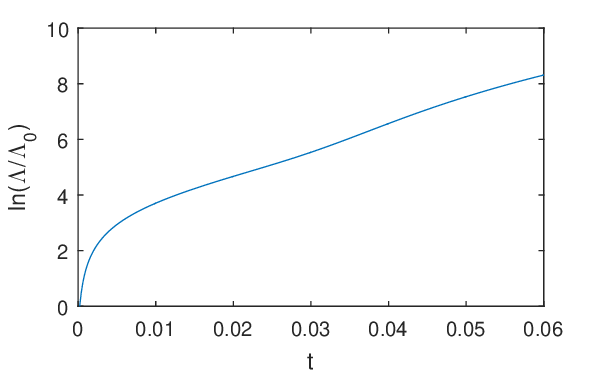}}
\caption{Super fast growth of the perturbation under different base solutions in 3D with parameters (\ref{3dfpe}).}
\label{3dfpf}
\end{figure}
Figure \ref{3dfpf} shows the super fast growth of the perturbation under different base solutions 
with initial conditions of the 
form (\ref{3dbi}), where $\La (t)$ is defined in (\ref{land}). 
We arrive at the same conclusion as in 2D, that is, the perturbation of higher mode base solutions has faster super fast growth, and such super fast growth is abundant among base solutions. Clearly, the super fast amplification of perturbations is a generic phenomenon that is independent of spatial dimensions, and is ubiquitous. 

\subsection{Turbulence Regime}

\begin{figure}[ht]
\centering
\subfigure[$N=16$]{\includegraphics[width=2.3in,height=2.3in]{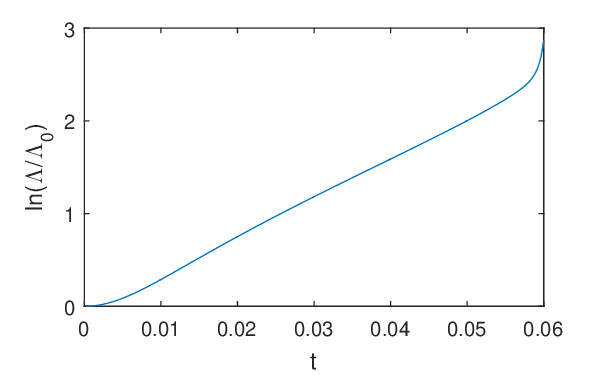}}
\subfigure[$N=8$]{\includegraphics[width=2.3in,height=2.3in]{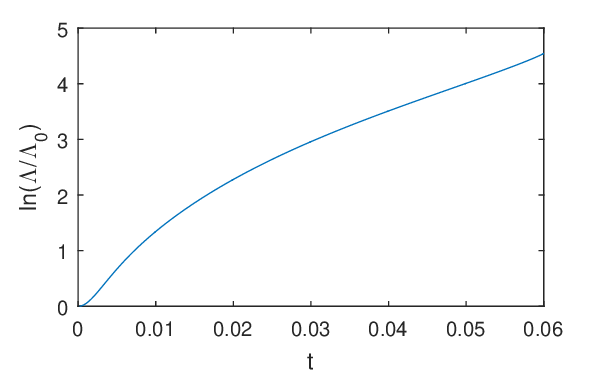}}
\subfigure[$N=4$]{\includegraphics[width=2.3in,height=2.3in]{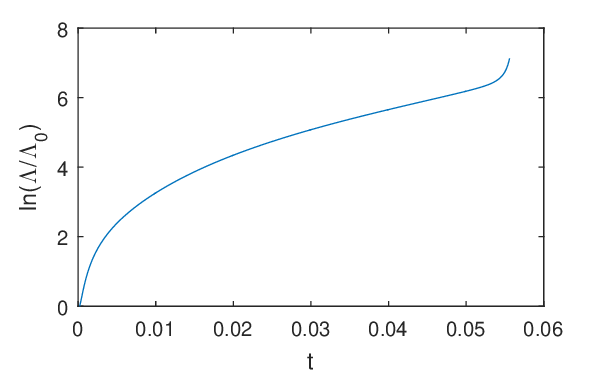}}
\subfigure[$N=2$]{\includegraphics[width=2.3in,height=2.3in]{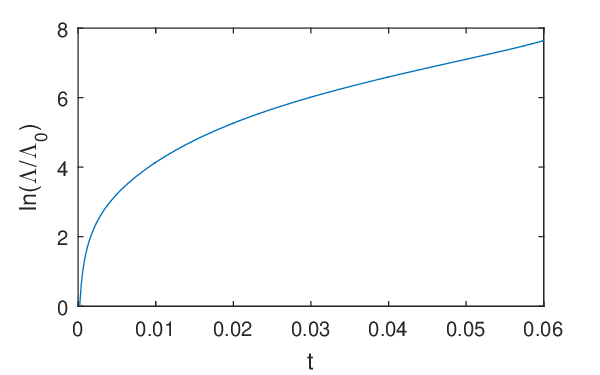}}
\caption{Super fast growths of the perturbations with the initial
condition (\ref{TIP3}) when $N=16, 8, 4, 2$, to the base solution with the initial condition (\ref{TIB3}),  where $\La (t) = \| du(t) \|_{H^3}$.}
\label{Tur3}
\end{figure}

In this subsection, we shall simulate more realistic situations of base solutions in the turbulence regime. We start with base solution's initial condition in the form
\begin{eqnarray}
u_1(0) &=& \sum_{l=1}^{16} \sum_{m=1}^{16} \sum_{n=1}^{16} a_{lmn} m n \cos (l x_1 + \phi_{1lmn}) 
\non \\ && \sin (m x_2+ \phi_{2lmn}) \sin (n x_3 + \phi_{3lmn}) , \non \\
u_2(0) &=& \sum_{l=1}^{16} \sum_{m=1}^{16} \sum_{n=1}^{16} a_{lmn} l n \sin (l x_1 + \phi_{1lmn}) \non \\ &&
\cos (m x_2+ \phi_{2lmn}) \sin (n x_3 + \phi_{3lmn}) , \label{TIB3} \\
u_3(0) &=& -2\sum_{l=1}^{16} \sum_{m=1}^{16} \sum_{n=1}^{16} a_{lmn} l m \sin (l x_1 + \phi_{1lmn}) \non \\ &&
\sin (m x_2+ \phi_{2lmn}) \cos (n x_3 + \phi_{3lmn}) , \non
\end{eqnarray}
and the initial condition for the perturbation in the form
\begin{eqnarray}
du_1(0) &=& \sum_{l=1}^{N} \sum_{m=1}^{N} \sum_{n=1}^{N} b_{lmn} m n \cos (l x_1 + \phi_{1lmn}) \non \\ &&
\sin (m x_2+ \phi_{2lmn}) \sin (n x_3 + \phi_{3lmn}) , \non \\
du_2(0) &=& \sum_{l=1}^{N} \sum_{m=1}^{N} \sum_{n=1}^{N} b_{lmn} l n \sin (l x_1 + \phi_{1lmn}) \non \\ &&
\cos (m x_2+ \phi_{2lmn}) \sin (n x_3 + \phi_{3lmn}) , \label{TIP3} \\
du_3(0) &=& -2\sum_{l=1}^{N} \sum_{m=1}^{N} \sum_{n=1}^{N} b_{lmn} l m \sin (l x_1 + \phi_{1lmn}) \non \\ &&
\sin (m x_2+ \phi_{2lmn}) \cos (n x_3 + \phi_{3lmn}) , \non
\end{eqnarray}
where $a_{lmn} = 0.0005 a$, $b_{lmn} = 0.00001 a$, $a$ is a random variable following the standard Gaussian
distribution, $\phi_{jlmn} = 2 \pi \phi$ ($j=1,2,3$), and $\phi$ is a random variable following the uniform
distribution over ($0,1$).
This type of initial conditions put the base flow into the turbulence regime.
We choose the Reynolds number $Re = 1000$, and we run the simulation
with time step $0.0001$. When $N=16, 8, 4, 2$, we have the same super fast growth (Fig. \ref{Tur3}), and clearly lower perturbation mode grows faster. We would like to reiterate that each initial individual Fourier mode amplifies independently. Finally, We would also like to reiterate that the super fast 
amplification phenomenon is a generic fact of fluids, and is ubiquitous. Microscopically, Navier-Stokes equations model fluid flows well. Thus, the super fast amplification phenomenon of Navier-Stokes equations reveals the same phenomenon in physical fluid flows.  

\section{Conclusion}

Through numerical simulations, we demonstrated the super fast growth of perturbations (rough dependence 
upon initial data) in high Reynolds number fluid flows. We also showed the abundance of such super fast growth among perturbations and base solutions in support of our theory that fully developed turbulence 
is caused and maintained by such super fast growth of perturbations. Such super fast 
amplification of perturbations is ubiquitous in turbulence.

\end{document}